# Empirical study on some interconnecting bilayer networks

Yan-Qin Qu, Xiu-Lian Xu, Shan Guan, Kai-Jun Li, Si-Jun Pan, Chang-Gui Gu, Yu-Mei Jiang, Da-Ren He*

*College of Physics Science and Technology, Yangzhou University, Yangzhou, 225002, P. R. China*

This manuscript serves as an online supplement of a preprint, which presents a study on a kind of bilayer networks where some nodes (called interconnecting nodes) in two layers merge. A model showing an important general property of the bilayer networks is proposed. Then the analytic discussion of the model is compared with empirical conclusions. We present all the empirical observations in this online supplement.

**1. Introduction**

In recent 10 years, scientists gradually accepted an idea that complex networks can be a good description tool for complex systems. However, most of the discussed networks only contain one kind of nodes and interactions. However, most of the complex systems contain many interdependent nodes and their interactions. We have to describe a complex system by many networks; each pair of them interacts also via some other networks [1,2]. Some scientists expressed the idea as *interdependent networks* [3,4], *supernetwork* [5], or *network of networks* [6-8].

Due to the difficulties for studying *supernetworks,* we may perform study on several interdependent network layers as the first step [1,2]. We present, in a preprint, an investigation on a kind of bilayer networks where some nodes appear in both layers. This makes the two layers interconnecting at the nodes. The common nodes, or "interconnecting nodes (IN)", play different roles in different layers. It is common to observe such interconnection of network layers in real world complex systems. For examples, some herbs (Chinese traditional medicines) may be simultaneously foods; they are IN of Chinese herb prescription and Chinese cooked dishes bilayer network (HP-CD bilayer) [9,10]. Some cities may contain both coach stations and railway stations; the cities are IN of coach traffic and train traffic bilayer network (coach-train bilayer) if we define cities as the nodes of the network layers (a possible definition of edge may be a direct traffic service between the neighboring two nodes).

If there are two network layers, $V_1=\{i_1,i_2,\cdots,i_{M1}\}$ denotes the node set of the first layer; $V_2=\{j_1,j_2,\cdots,j_{M2}\}$

denotes the node set of the second layer, and $V_1 \cap V_2 \neq \varphi$, we define the nodes in the nonempty intersection as IN, and the bilayer system as an "interconnecting bilayer network".

Many network statistical topological properties have been investigated [11]. Some of them are suitable for a description of "averaged differences of IN topological properties (ADTP)". In our opinion, node degree, which is defined as the node neighboring edge number, should be the most suitable one. Betweenness, which is defined as the shortest path number passing through the node, is the second choice. Let $x$ denote one of the properties, we define "differences of single IN network topological properties (DSTP)" of an IN described by $x$ as

$$u_x = \left| \frac{x_i}{\langle x_i \rangle} - \frac{x_j}{\langle x_j \rangle} \right|, \tag{1}$$

where $x_i$ represents $x$ value of an IN, $i$, in the upper layer, $x_j$ represents $x$ value of the same IN, but with a sequence number $j$ in the lower layer, $\langle x_i \rangle$ denotes the averaged $x_i$ value of all the nodes in the upper layer, $\langle x_j \rangle$ that in the lower layer. Accordingly, we define ADTP of the bilayer described by $x$ as

$$U_x = \frac{1}{m} \sum_1^m \left| \frac{x_i}{\langle x_i \rangle} - \frac{x_j}{\langle x_j \rangle} \right|, \tag{2}$$

where $m$ denotes "interconnecting node number (INN)". The normalized INN (NINN) is defined as

$$n = \frac{2m}{M_i + M_j}, \tag{3}$$

where $M_i$ denotes the total node number of the upper layer, $M_j$ that of the lower layer.

In the preprint [12], we present a model showing an important property, the general relation between $U_x$ and $n$, of the bilayer networks. The analytic discussion of the model then is compared with empirical conclusions. The empirical investigation statements are quite long and boring, therefore, we present all the empirical observations in this online supplement.

**2. Empirical Investigation**

We empirically investigated distributions, $P(x)$, of degree and betweenness, $U_x$ (including $U_k$, ADTP described by degree, and $U_b$, ADTP described by betweenness), $n$ (NINN), and the relation between $u_x$ and the averaged property value $x_a=(x_i+x_j)/2$ (including that for degree, $k_a=(k_i+k_j)/2$ and for betweenness, $b_a=(b_i+b_j)/2$) in eight real world bilayer networks.

All the $P(x)$ distribution data can be fitted by function $P(x) \propto (x+\alpha)^{-\gamma}$ [10, 13]. If $\alpha=0$, it becomes a power law; it approaches an exponential function when $\alpha \rightarrow \infty$. So, the function interpolates between a power law and an exponential decay. We can characterize each property distribution by two parameters, $\alpha$ and $\gamma$. In the following, the real world bilayer networks are labeled from 1 through 8 in increasing order of $n$, i.e. $n_1 \leq n_2 \leq \cdots \leq n_8$.

## 2.1 Empirical Investigation on Chinese herb prescription-Chinese cooked dishes bilayer network (HP-CD)

**Bilayer network 1:** In the first layer, we define herbs as nodes; two nodes are connected by an edge if they appear in at least one common prescription [10]. The data contain 917 prescriptions and 1612 herbs. There are 23035 edges between the nodes. The data were collected from a book [14], which collects the main ancient and present herb prescriptions. In the second layer, foods are defined as nodes; two nodes are connected if they appear in at least one common cooked dish [9]. 534 cooked dishes and 595 foods were included in the data, which were collected from another book [15]. There are 7876 edges between the nodes. The book collects the famous family applicable cooked dishes. An IN simultaneously is an herb and a food. 43 INs were found in the data.

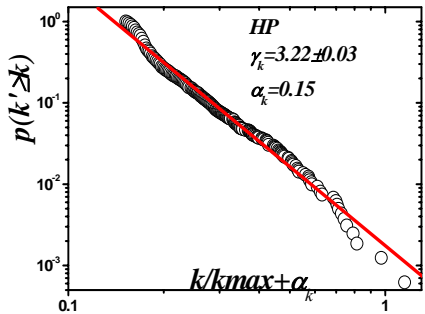 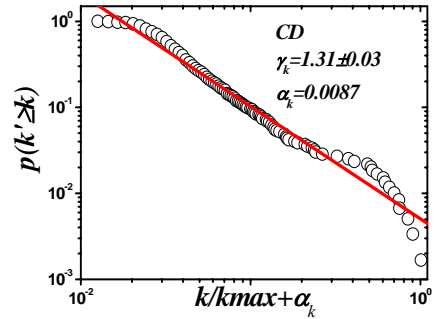

Fig. 2.1.1 Cumulative distribution of degree in herb layer.　　Fig. 2.1.2 Cumulative distribution of degree in food layer.

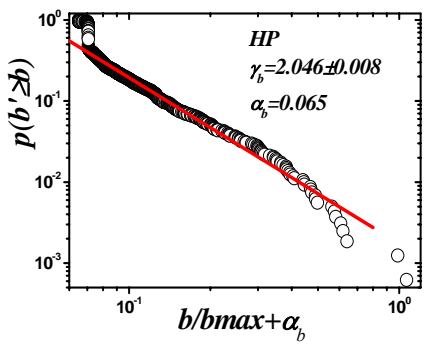 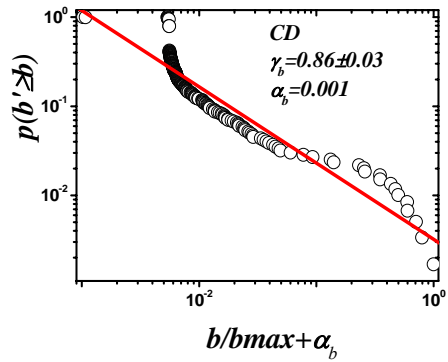

Fig. 2.1.3 Cumulative distribution of betweenness in herb layer.　　Fig. 2.1.4 Cumulative distribution of betweenness in food layer.

We show cumulative distributions of degree and betweenness in both the layers by four figures (Fig. 2.1.1-2.1.4). Degree and betweenness are normalized by their maximum values, which are labeled in the figures by *kmax* and *bmax*, respectively. As can be seen, all the distributions can be fitted by SPL functions.

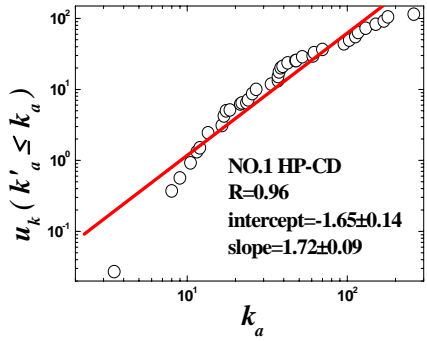 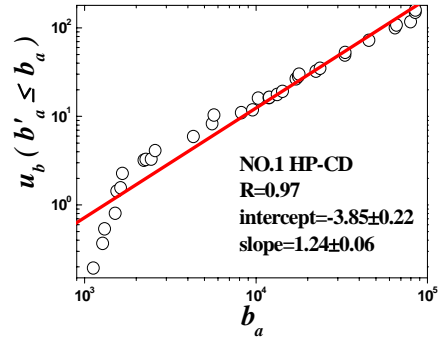

Fig. 2.1.5 Relation between $u_k$ and $k_a$ in system 1.   Fig. 2.1.6 Relation between $u_b$ and $b_a$ in system 1.

Figures 2.1.5 and 2.1.6 present the cumulative relation between $u_x$ and the averaged property value $x_a=(x_i+x_j)/2$ (including that for degree, $k_a=(k_i+k_j)/2$ and for betweenness, $b_a=(b_i+b_j)/2$) in this bilayer. They can be approximately fitted by power law functions.

## 2.2 Empirical Investigation on Biology keyword - Physics keyword bilayer network (BK-PK)

**Bilayer network 2:** We define key words as nodes in both the layers. Two nodes are connected by an edge if they appear in at least one common scientific paper. The data contain 1416 biology papers and 4495 biology key words. There are 11183 edges between the nodes. The data contain 1037 physics papers and 3028 physics key words. There are 5166 edges between the nodes. An IN simultaneously is a biology key word and a physics key word. 169 INs were found in the data, which were downloaded from http://wulixb.iphy.ac.cn/cn/ch/index.aspx and http://lunwen.cnetnews.com.cn/. The data include the key words of all the papers published in a famous Chinese physics journal, "Acta Physica Sinica", in 2005 and the key words of all the Chinese biology conference papers published in 2005.

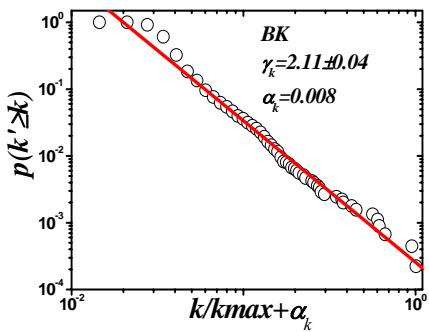 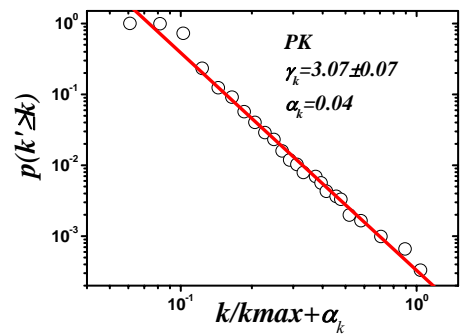

Fig. 2.2.1 Cumulative distribution of degree in biology layer.   Fig. 2.2.2 Cumulative distribution of degree in physics layer.

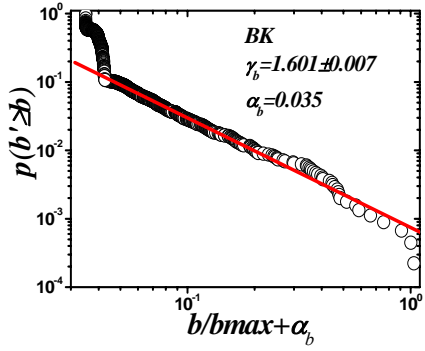 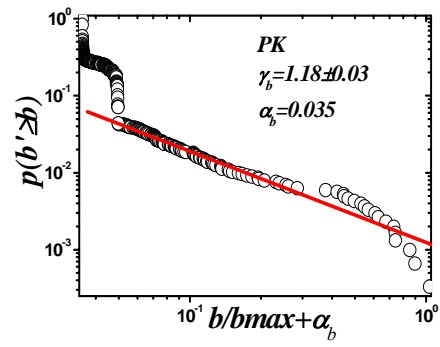

Fig. 2.2.3 Cumulative distribution of betweenness in biology layer. Fig. 2.2.4 Cumulative distribution of betweenness in physics layer.

We show cumulative distributions of degree and betweenness in both the layers by figures 2.2.1-2.2.4. Degree and betweenness are normalized by their maximum values. All the distributions can be fitted by SPL functions.

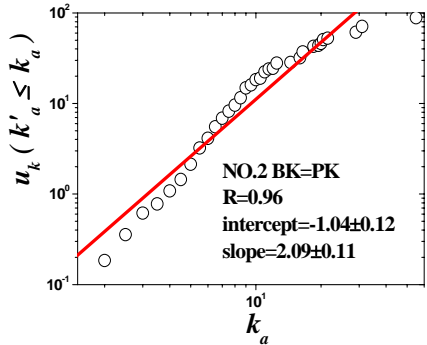 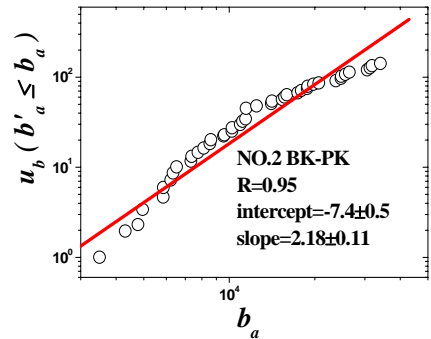

Fig. 2.2.5 Relation between $u_k$ and $k_a$ in system 2.    Fig. 2.2.6 Relation between $u_b$ and $b_a$ in system 2.

Figures 2.2.5 and 2.2.6 present the cumulative relation between $u_x$ and the averaged property value $x_a=(x_i+x_j)/2$ (including that for degree, $k_a=(k_i+k_j)/2$ and for betweenness, $b_a=(b_i+b_j)/2$) in this bilayer. They can be approximately fitted by power law functions.

## 2.3 Empirical Investigation on Chinese mainland movie actor - Chinese Hong Kong movie actor bilayer network (MMA-HKMA)

**Bilayer network 3:** We define movie actors as nodes in both the layers. Two nodes are connected by an edge in the first layer if they perform in at least one common mainland company movie or in the second layer if they perform in at least one common Hong Kong company movie in 2005 and 2006. The data contain 869 mainland movies and 3219 mainland movie actors. There are 44153 edges between the nodes. The data contain 337 Hong

Kong movies and 1132 Hong Kong movie actors. There are 11455 edges between the nodes. An IN performs in both mainland movies and Hong Kong movies in the two years. 361 INs were found in the data, which were downloaded from http://www.cnmdb.com and http://mdbchina.com.

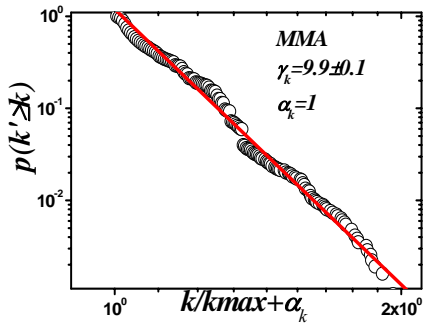 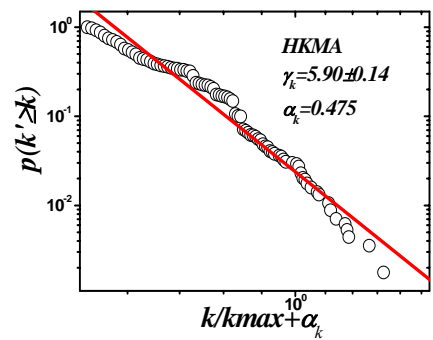

Fig. 2.3.1 Cumulative distribution of degree in mainland layer.   Fig. 2.3.2 Cumulative distribution of degree in HK layer.

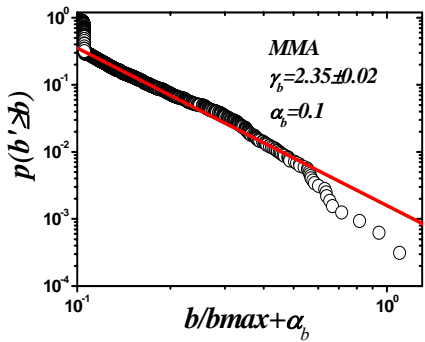 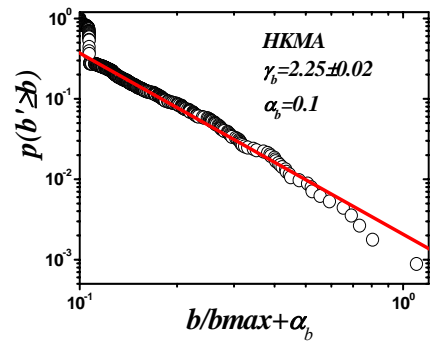

Fig. 2.3.3 Cumulative distribution of betweenness in mainland layer. Fig. 2.3.4 Cumulative distribution of betweenness in HK layer.

We show cumulative distributions of degree and betweenness in both the layers by figures 2.3.1-2.3.4. Degree and betweenness are normalized by their maximum values. All the distributions can be fitted by SPL functions.

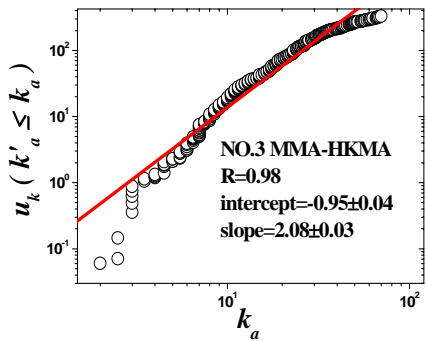 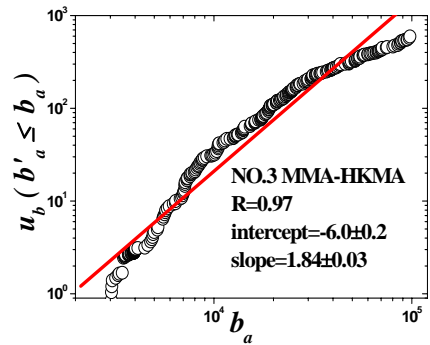

Fig. 2.3.5 Relation between $u_k$ and $k_a$ in system 3.    Fig. 2.3.6 Relation between $u_b$ and $b_a$ in system 3.

Figures 2.3.5 and 2.3.6 present the cumulative relation between $u_x$ and the averaged property value $x_a=(x_i+x_j)/2$ (including that for degree, $k_a=(k_i+k_j)/2$ and for betweenness, $b_a=(b_i+b_j)/2$) in this bilayer. They can be approximately fitted by power law functions.

## 2.4 Empirical Investigation on Baker's yeast protein interaction - metabolic bilayer network (YPI-YM)

**Bilayer network 4:** We define proteins as nodes and physical interactions between proteins as edges in the first layer [16,17]. The data contain 3985 proteins and 30677 edges. We define enzymes (a kind of proteins) as nodes and common biochemical compounds shared by two enzymatic reactions as edges in the second layer [18]. The data contain 527 enzymes and 38285 edges. An IN is defined if both protein interaction and metabolic layers share the protein. 445 INs were found in the data, which were downloaded from http://www.ebi.ac.uk/intact/ and http://www.genome.jp/.

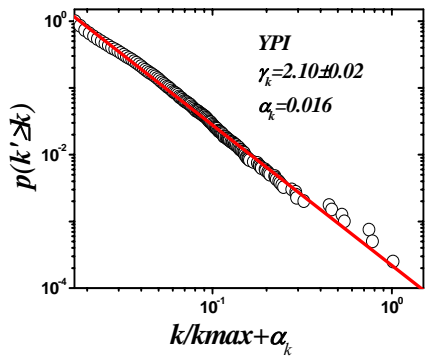
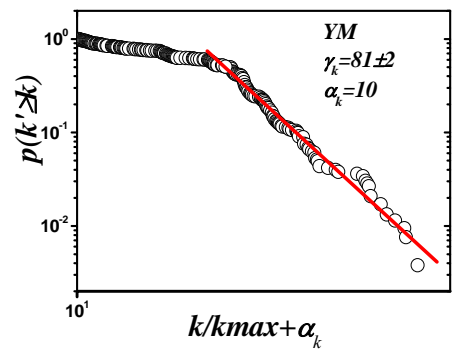

Fig. 2.4.1 Cumulative distribution of degree in YPI layer.   Fig. 2.4.2 Cumulative distribution of degree in YM layer.

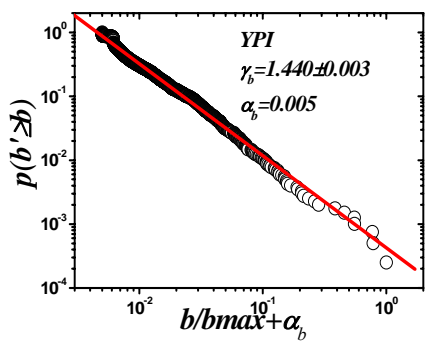
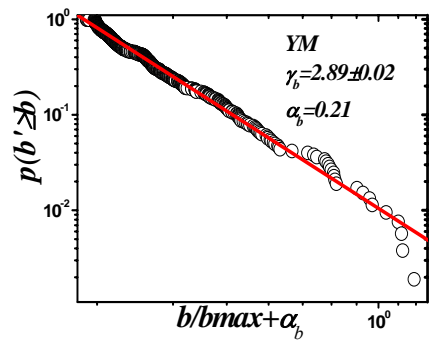

Fig. 2.4.3 Cumulative distribution of betweenness in YPI layer.   Fig. 2.4.4 Cumulative distribution of betweenness in YM layer.

We show cumulative distributions of degree and betweenness in both the layers by figures 2.4.1-2.4.4. Degree and betweenness are normalized by their maximum values. All the distributions can be fitted by SPL functions.

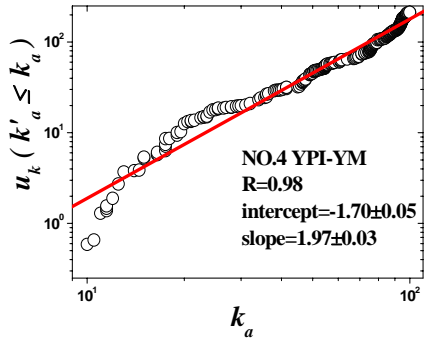 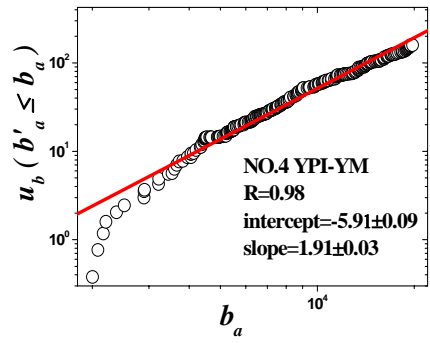

Fig. 2.4.5 Relation between $u_k$ and $k_a$ in system 4.　　Fig. 2.4.6 Relation between $u_b$ and $b_a$ in system 4.

Figures 2.4.5 and 2.4.6 present the cumulative relation between $u_x$ and the averaged property value $x_a=(x_i+x_j)/2$ (including that for degree, $k_a=(k_i+k_j)/2$ and for betweenness, $b_a=(b_i+b_j)/2$) in this bilayer. They can be approximately fitted by power law functions.

## 2.5 Empirical Investigation on E.coli-K12 protein interaction - metabolic bilayer network (EPI-EM)

**Bilayer network 5:** The definitions and data sources are the same as in Baker's yeast protein interaction - metabolic bilayer network [15,16,17]. The data contain 2893 proteins, and 14009 edges in the first layer and 758 enzymes and 63035 edges in the second layer. 623 INs were found in the data

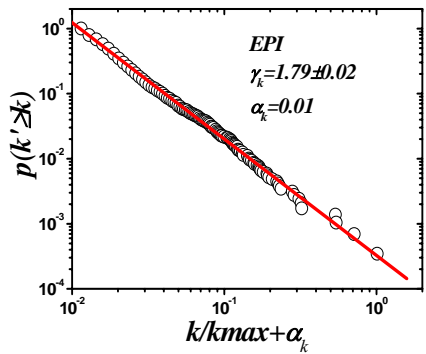 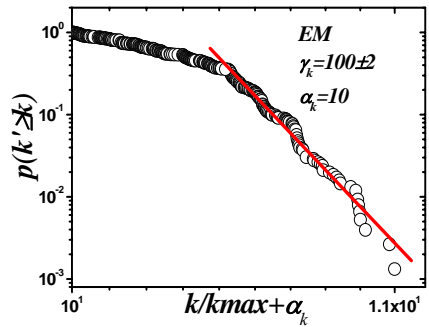

Fig. 2.5.1 Cumulative distribution of degree in EPI layer.　　Fig. 2.5.2 Cumulative distribution of degree in EM layer.

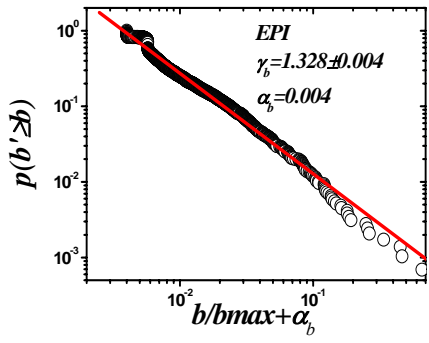 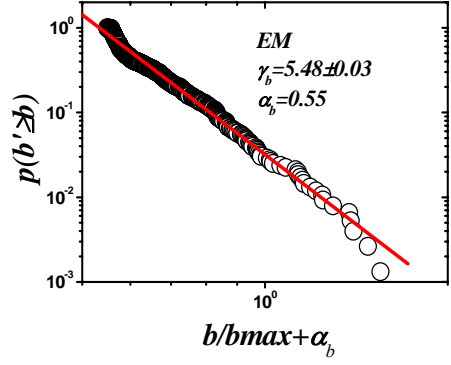

Fig. 2.5.3 Cumulative distribution of betweenness in EPI layer.    Fig. 2.5.4 Cumulative distribution of betweenness in EM layer.

We show cumulative distributions of degree and betweenness in both the layers by figures 2.5.1-2.5.4. Degree and betweenness are normalized by their maximum values. All the distributions can be fitted by SPL functions.

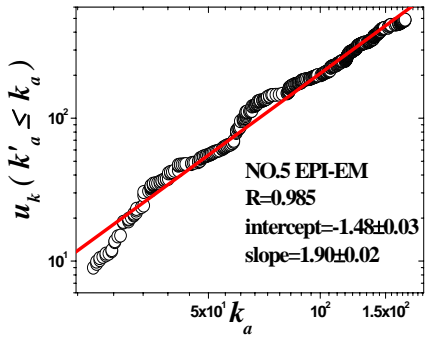 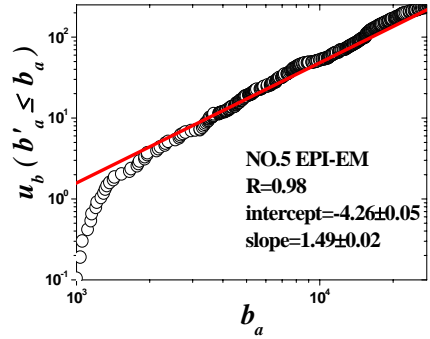

Fig. 2.5.5 Relation between $u_k$ and $k_a$ in system 5.    Fig. 2.5.6 Relation between $u_b$ and $b_a$ in system 5.

Figures 2.5.5 and 2.5.6 present the cumulative relation between $u_x$ and the averaged property value $x_a=(x_i+x_j)/2$ (including that for degree, $k_a=(k_i+k_j)/2$ and for betweenness, $b_a=(b_i+b_j)/2$) in this bilayer. They can be approximately fitted by power law functions.

**2.6 Empirical Investigation on Chinese Traffic bilayer networks:**

**Bilayer network 6: coach - airplane bilayer network (coach - airplane),**

**Bilayer network 7: train – airplane bilayer network (train – airplane),**

**Bilayer network 8: coach – train bilayer network (coach –train)**

We define cities containing coach stations, train stations, or airports with an administration level higher than "regional counties" (China mainland is divided into 31 provinces, which are further divided into 333 regional counties) as nodes. Two nodes are connected by an edge if corresponding traffic tool (coach, train or airplane) provides direct traffic service (without changing coach, train or airplane) between them. The data, which were downloaded from http://www.china-holiday.com, http://www.ipao.com, http://train.hepost.com/, contain 314 cities with coach stations and 3220 edges in the coach layer, 251 cities with train stations and 6775 edges in the train layer, and 100 cities with airports and 838 edges in the airplane layer. An IN is defined if a city contains two kinds of stations, i.e., a coach station and an airport, a train station and an airport, or a coach station and a train station. 100 INs were found in the coach-airplane bilayer data. 88 INs were found in the train–airplane bilayer data. 251 INs were found in the coach- train bilayer data.

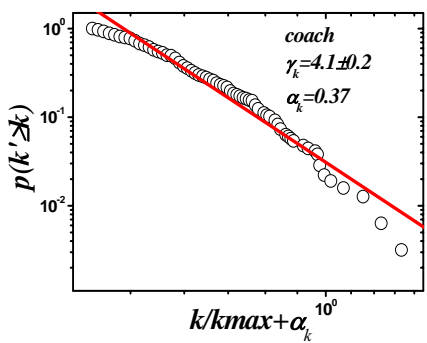
Fig. 2.6.1 Cumulative distribution of degree in coach layer.

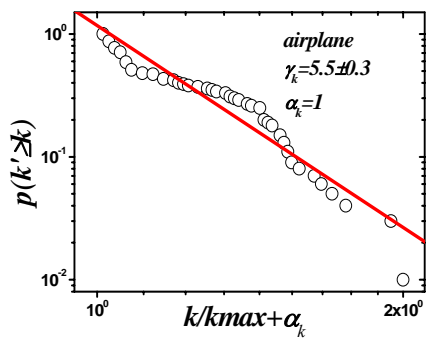
Fig. 2.6.2 Cumulative distribution of degree in airplane layer.

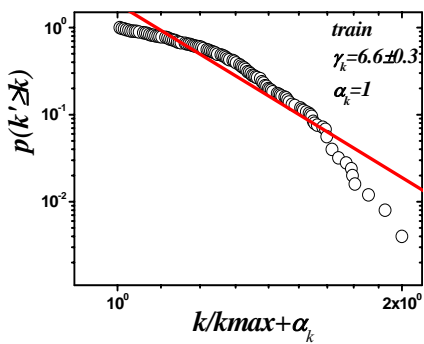
Fig. 2.6.3 Cumulative distribution of degree in train layer.

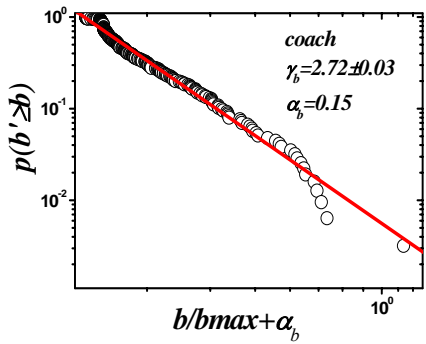
Fig. 2.6.4 Cumulative distribution of betweenness in coach layer.

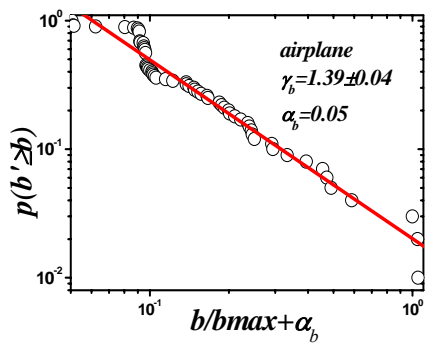
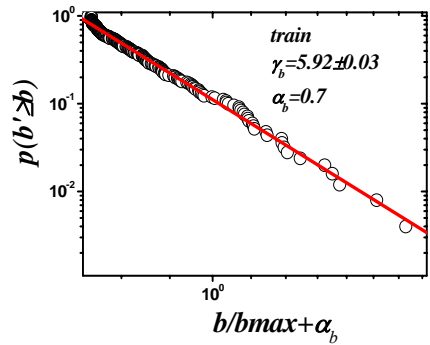

Fig. 2.6.5 Cumulative distribution of betweenness in airplane layer.  Fig. 2.6.6 Cumulative distribution of betweenness in train layer.

We show cumulative distributions of degree and betweenness in the three layers by figures 2.6.1-2.6.6. Degree and betweenness are normalized by their maximum values. All the distributions can be fitted by SPL functions.

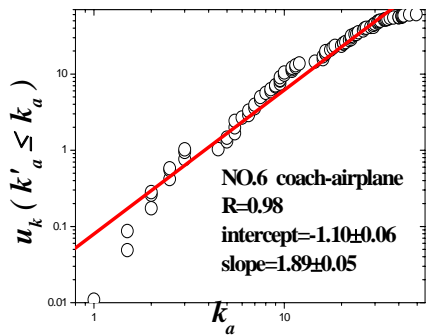
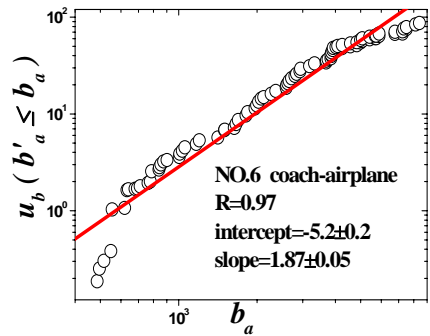

Fig. 2.6.7 Relation between $u_k$ and $k_a$ in system 6.  Fig. 2.6.8 Relation between $u_b$ and $b_a$ in system 6.

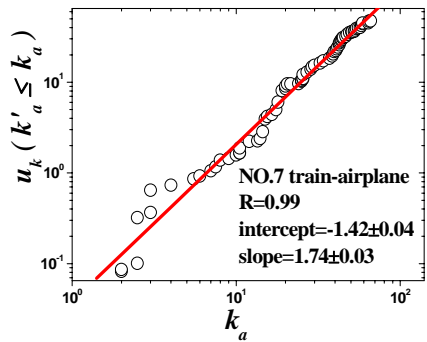
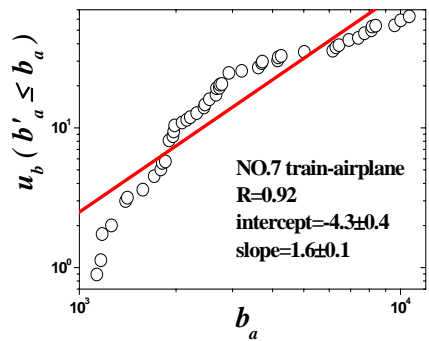

Fig. 2.6.9 Relation between $u_k$ and $k_a$ in system 7.  Fig. 2.6.10 Relation between $u_b$ and $b_a$ in system 7.

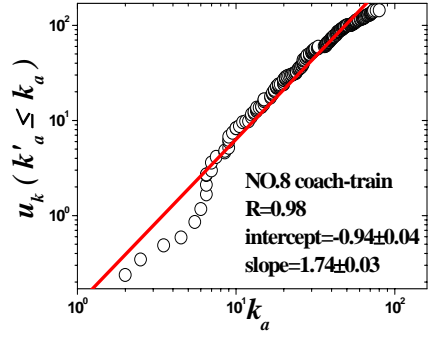 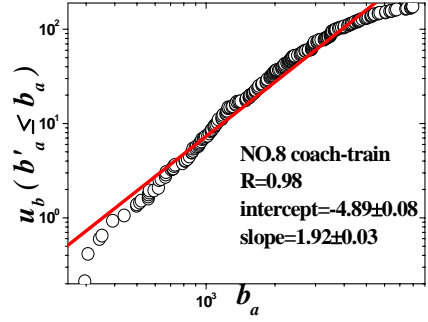

Fig. 2.6.11 Relation between $u_k$ and $k_a$ in system 8.     Fig. 2.6.12 Relation between $u_b$ and $b_a$ in system 8.

Figures 2.6.7-2.6.12 present the cumulative relation between $u_x$ and the averaged property value $x_a=(x_i+x_j)/2$ (including that for degree, $k_a=(k_i+k_j)/2$ and for betweenness, $b_a=(b_i+b_j)/2$) in bilayer 6-8. They can be approximately fitted by power law functions.

Table 1: Parameters of the eight bilayer networks. $M$ and $e$ are the node number and edge number of each network bilayer, respectively. $m$ and $n$ denote the number and the normalized number of the interconnecting nodes, respectively. $\alpha_k$, $\gamma_k$, $\alpha_b$, $\gamma_b$ denote the fitted SPL parameters of the degree distribution and betweenness distribution. $U_k$ and $U_b$ denote the average differences of the node degree and betweenness of the interconnecting nodes, respectively. $\gamma_{uk-ka}$ and $\gamma_{ub-ba\ b}$ denote the fitted power exponents of the correlation between the $u_k$ and $k_a$, and between the $u_b$ and $b_a$, respectively. The details of deriving these parameters are presented in Ref. [12].

| Bilayer No. | Layer | $M$ | $e$ | $\alpha_k$ | $\gamma_k$ | $\alpha_b$ | $\gamma_b$ | $m$ | $n$ | $U_k$ | $U_b$ | $\gamma_{uk-ka}$ | $\gamma_{ub-ba}$ |
|---|---|---|---|---|---|---|---|---|---|---|---|---|---|
| 1 | HP | 1612 | 23035 | 0.15 | 3.22 | 0.065 | 2.046 | 43 | 0.039 | 2.69 | 3.76 | 0.72 | 0.24 |
| 1 | CD | 595 | 7876 | 0.0087 | 1.31 | 0.001 | 0.86 | | | | | | |
| 2 | BK | 4495 | 11183 | 0.008 | 2.11 | 0.035 | 1.601 | 169 | 0.045 | 1.20 | 3.41 | 1.09 | 1.18 |
| 2 | PK | 3028 | 5166 | 0.04 | 3.07 | 0.035 | 1.18 | | | | | | |
| 3 | MMA | 3219 | 44153 | 1 | 9.9 | 0.1 | 2.35 | 361 | 0.17 | 1.04 | 2.88 | 1.08 | 0.84 |
| 3 | HKMA | 1132 | 11455 | 0.475 | 5.9 | 0.1 | 2.25 | | | | | | |
| 4 | YPI | 3985 | 30677 | 0.016 | 2.10 | 0.005 | 1.440 | 445 | 0.20 | 0.97 | 1.32 | 0.97 | 0.91 |
| 4 | YM | 527 | 38285 | 10 | 81 | 0.21 | 2.89 | | | | | | |
| 5 | EPI | 2893 | 14009 | 0.01 | 1.79 | 0.004 | 1.328 | 623 | 0.34 | 1.03 | 1.29 | 0.90 | 0.49 |
| 5 | EM | 758 | 63035 | 10 | 100 | 0.55 | 5.48 | | | | | | |
| 6 | coach | 314 | 3220 | 0.37 | 4.1 | 0.15 | 2.72 | 100 | 0.48 | 0.72 | 1.07 | 0.89 | 0.87 |
| 6 | airplane | 100 | 838 | 1 | 5.5 | 0.05 | 1.39 | | | | | | |
| 7 | train | 251 | 6775 | 1 | 6.6 | 0.7 | 5.92 | 88 | 0.50 | 0.66 | 0.91 | 0.74 | 0.6 |
| 7 | airplane | 100 | 838 | 1 | 5.5 | 0.05 | 1.39 | | | | | | |
| 8 | coach | 314 | 3220 | 0.37 | 4.1 | 0.15 | 2.72 | 251 | 0.89 | 0.61 | 0.89 | 0.74 | 0.92 |
| 8 | train | 251 | 6775 | 1 | 6.6 | 0.7 | 5.92 | | | | | | |

Table 1 summarizes properties and parameters of all the bilayer networks.

## 3. The model and discussions

The empirical observation results show that most of the $u_x$-$x_a$ data can be fitted approximately by power law functions with the scaling exponents around 1.0. The empirical results suggest a relation, $\lg u_x = (1 \pm \delta(l))\lg x_a + \nu(l)$, where $\delta(l)$ is a small quantity. Ignoring $\delta(l)$, we get $u_x \cong \eta(l)x_a$ where $\eta(l)=10^{\nu(l)}$ is a proportional coefficient depending on bilayer system index $l$. Accordingly, we have $(x_i-x_j)/<x>=\eta(l)x_a=\eta(l)(x_i+x_j)/2$, which immediately lead to $x_i=\xi(l)x_j$ with $\xi(l)=[1+\eta(l)<x>/2]/[1-\eta(l)<x>/2]>1$ depending on bilayer system index $l$. A very simple model was proposed in [12] based on the empirical observations.

Although we have only 16 $U_x$-$n$ data (eight for $U_k$ and eight for $U_b$), a comparison of the model analysis with these data is still meaningful. In [12] we present a comparison of the model analysis with the 16 empirical data. The comparison shows that, with such a simple model, the analytic conclusion still show rather good agreement with the empirical results. This shows that the model describes key point of the bilayer network characteristics.

In conclusion, we propose a very simple and straightforward model describing a new kind of interdependence (merging of some nodes) of two "network layers". The merge represents that the interconnecting nodes play different roles in both the layers. The analytic treatment of the model gives rise to a relation function between the interconnecting node number and the averaged differences of interconnecting node degrees and betweenness. The result shows rather good agreement with the empirical conclusions obtained in eight real world bilayer networks.

This work is supported by the Chinese National Natural Science Foundation under grant numbers 10635040 and 70671089.

* E- address: darendo10@yahoo.com.cn